\documentclass[aps,prl,twocolumn,floatfix,reprint,superscriptaddress]{revtex4}
\usepackage{eurosym}
\usepackage{amsfonts}
\usepackage{mathrsfs}
\usepackage{amsmath}
\usepackage{color}
\usepackage{graphicx}
\usepackage{bm}
\usepackage{amssymb}
\usepackage{xspace}
\usepackage{epstopdf}
\usepackage{dcolumn}
\usepackage{tabularx}
\usepackage{longtable}
\usepackage{mhchem}
\usepackage[colorlinks=true, letterpaper=true, pdfstartview=FitV, linkcolor=blue, citecolor=blue, urlcolor=blue]{hyperref}
\usepackage[normalem]{ulem}

\begin{document}

\title{Topological Mott Insulator with Bosonic Edge Modes in 1D Fermionic
Superlattices}
\author{Haiping Hu}
\affiliation{Department of Physics, The University of Texas at Dallas, Richardson, Texas
75080, USA}
\author{Shu Chen}
\affiliation{Beijing National Laboratory for Condensed Matter Physics, Institute of
Physics, Chinese Academy of Sciences, Beijing 100190, China}
\affiliation{School of Physical Sciences, University of Chinese Academy of Sciences,
Beijing, 100049, China}
\author{Tian-Sheng Zeng}
\affiliation{Department of Physics, The University of Texas at Dallas, Richardson, Texas
75080, USA}
\author{Chuanwei Zhang}
\email{chuanwei.zhang@utdallas.edu}
\affiliation{Department of Physics, The University of Texas at Dallas, Richardson, Texas
75080, USA}

\begin{abstract}
We investigate topological phase transitions driven by interaction and
identify a novel topological Mott insulator state in one-dimensional
fermionic optical superlattices through numerical density matrix
renormalization group (DMRG) method. Remarkably, the low-energy edge
excitations change from spin-1/2 \textit{fermionic} single-particle modes to
spin-1 \textit{bosonic} collective modes across the phase transition. Due to
spin-charge separation, the low-energy theory is governed by an effective
spin superexchange model, whereas the charge degree of freedom is fully
gapped out. Such topological Mott state can be characterized by a spin Chern
number and gapless magnon modes protected by a finite spin gap. The proposed
experimental setup is simple and may pave the way for the experimental
observation of exotic topological Mott states.
\end{abstract}

\maketitle

\textit{Introduction.---}The interplay between single-particle band topology
and many-body interaction plays a crucial role in many important
strongly-correlated phenomena in condensed matter physics. Unlike
single-particle topological states \cite{review1,review2}, interactions can
induce remarkable physical phenomena such as fractionalization of emergent
collective excitations and give rise to intriguing correlated states that
exhibit non-trivial topological properties. Prominent examples are
fractional quantum Hall effects \cite{fractional1,fractional2} where
constituent particles are electrons but emergent quasiparticles only carry
fractions of electron charge, and topological Mott insulator \cite{topomott1}
with deconfined spinon excitations.

One-dimensional (1D) interacting systems, which are amenable to exact
methods, provide fundamental insights for understanding strongly-correlated
states. Due to the confined geometry, the low-energy excitations are
collective and exhibit a peculiar fractionalization, spin-charge separation
(SCS). A single-particle excitation is divided into two collective modes,
which possess charge and spin degrees of freedom, respectively. In a
topological Mott insulator, low-energy excitations lie in the spin sector,
whereas charge excitations are frozen by strong interactions. The
topological properties manifest themselves by the appearance of gapless
modes at the boundary protected by the insulating bulk. In previous studies,
these edge modes are composed of spinons~\cite{topomott2,topomott3} carrying
spin-1/2 and no charge. As spinful modes can also carry integer spin (like
magnon, spin-1), two natural and important questions need to be addressed:
i) Are there topological Mott insulator states hosting other types of
spinful edge modes and how to characterize them? ii) Since SCS severely
changes the low-energy excitations, what is the bulk-edge correspondence in
a topological Mott insulator state?

In this paper, we address these two important questions by studying the Mott
insulator states in a 1D fermionic optical superlattice. Ultracold atoms in
optical lattices have provided unprecedented controllability to simulate
strongly interacting systems. In particular, 1D optical superlattices open a
new and simple avenue towards realizing exotic topological states \cite%
{lang,kraus,shiliangzhu,xuzhihao,xuzhihaotopomott,hhpmag1,hhpmag2} because
of their exact correspondence with quantum Hall physics \cite%
{hofstadter,harper} in an extended space. Here we scrutinize Mott
transitions in 1D optical superlattices and identify a novel topological
Mott insulator using the quasi-exact numerical density matrix
renormalization group (DMRG). Such Mott phase transition from a band
topological insulator to a topological Mott insulator is accompanied by bulk
excitation gap closing, SCS, and the change of topological spin Chern
number. The corresponding low-energy excitations change from single-particle
spin-1/2 fermionic modes to collective spin-1 bosonic modes, in consistent
with the spin Chern number change across the transition (\textit{i.e.},
bulk-edge correspondence). The low-energy physics is governed by an
antiferromagnetic spin superexchange model due to SCS. Our proposed
experimental setup involves fermions in a 1D triple-well superlattice and is
simple to realize in experiments comparing to other complex lattice models
or materials~\cite{topomott1,topomott2,topomott3}.

\begin{figure}[t]
\centering
\includegraphics[width=3.1in]{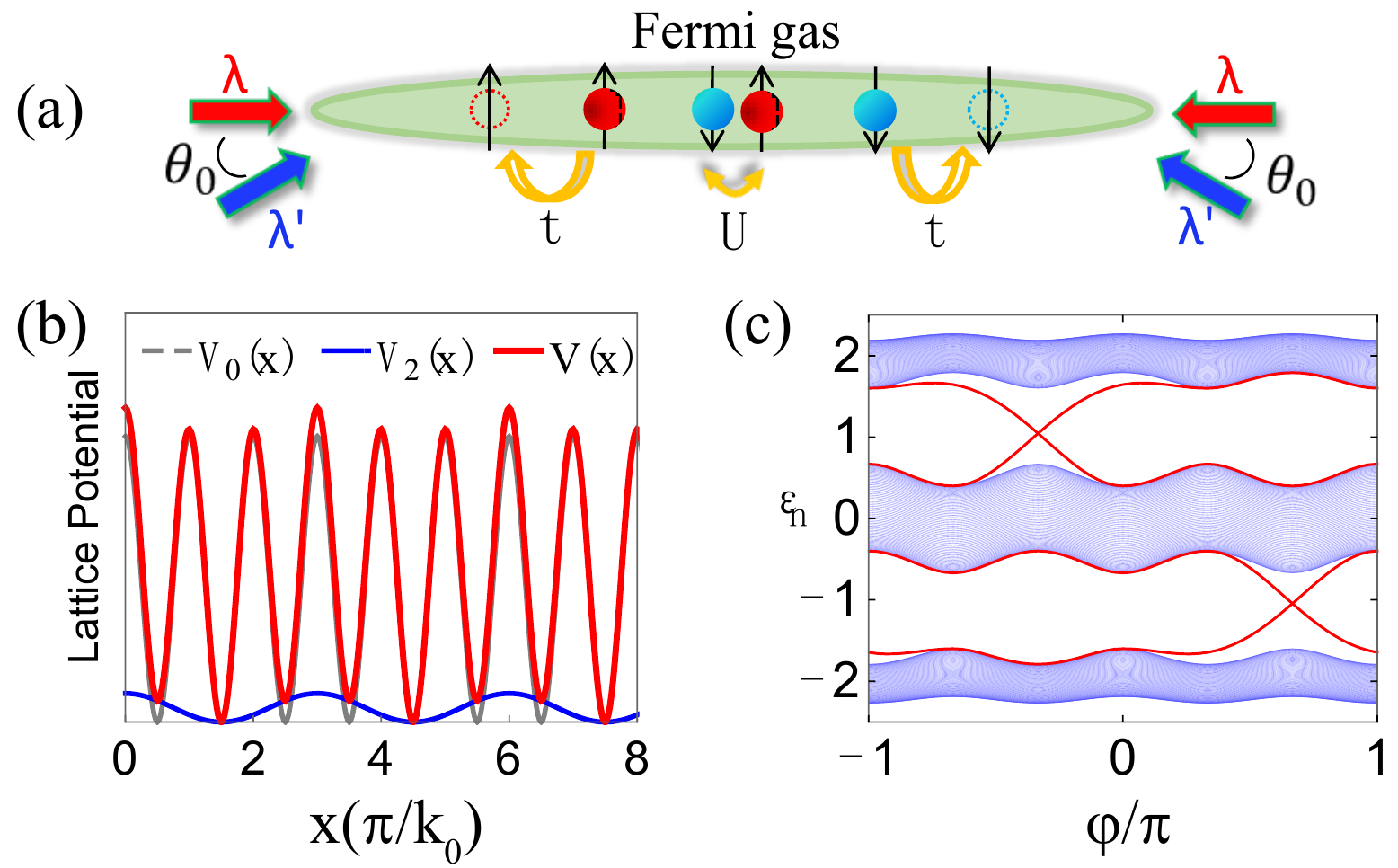}
\caption{(a)~Schemetic of experimental setup. A two-component Fermi gas is
confined in 1D. Two counter-propagating red lasers (wavelength $\protect%
\lambda $) form the main lattice, while two blues lasers ($\protect\lambda %
^{\prime }$) incident at an angle $\protect\theta _{0}$ form the secondary
weak long period lattice. (b)~Resulting lattice potential. The local minima
of the total potential are located at the minima of the main lattice.
(c)~Single-particle bands under OBC. Red lines highlight the end modes. $q=3$%
, $\protect\mu =1.2$, $L=240$.}
\label{fig1}
\end{figure}

\textit{Single-particle physics.---}We consider 1D Fermi gases with two
internal states (labeled by $\sigma =\uparrow ,\downarrow $) tightly
confined in transverse directions [Fig. \ref{fig1}(a)]. Two
counter-propagating laser beams (wavelength $\lambda $) form a main optical
lattice $V_{0}(x)=V_{0}\cos ^{2}(k_{0}x)$ with $k_{0}=2\pi /\lambda $. Two
additional laser beams (wavelength $\lambda ^{\prime }$) incident at a tilt
angle $\theta _{0}$ form a secondary weak lattice $V_{2}(x)=V_{2}\cos
^{2}(k_{2}x+\varphi )$ with $k_{2}=2\pi \cos \theta _{0}/\lambda ^{\prime }$
and relative phase $\varphi $ with respect to the main lattice. As
illustrated in Fig. \ref{fig1}(b), the total potential $%
V(x)=V_{0}(x)+V_{2}(x)$ forms a superlattice, with its period determined by
the ratio $q=k_{0}/k_{2}=\frac{\lambda ^{\prime }}{\lambda \cos \theta _{0}}$%
. Such optical superlattice has been experimentally realized by many groups~%
\cite{potential1,potential2,potential3}.

When the potential depth $V_{0}$ is much larger than the recoil energy $%
E_{r}=\hbar ^{2}k_{0}^{2}/2M$ ($M$ is the atomic mass), only the lowest
Bloch band need be considered and the system can be well approximated \cite%
{superlattice,giamarchi,superlattice2,superlattice3} by the following
tight-binding model with a superlattice potential
\begin{equation}
H=\sum_{j=1,\sigma }^{L}[-t(c_{j\sigma }^{\dag }c_{j+1\sigma }+h.c.)+\mu
_{j}n_{j\sigma }]+Un_{j\uparrow }n_{j\downarrow },  \label{ham}
\end{equation}%
where $c_{j\sigma }$ annihilates a spin-$\sigma $ fermion at $j$-th site, $L$
is the length of the chain, and $\mu _{j}=\mu \cos (2\pi j/q+\varphi )$ is
the on-site long period superlattice potential. Hereafter $t=1$ is set as
the energy unit. A cyclical variation of $\varphi $ brings back the
Hamiltonian $H(\varphi )=H(\varphi +2\pi )$, which is crucial for
topological Thouless pumping \cite{thouless} and the evolution of edge
modes. The interaction $U$ between atoms can be tuned over a wide range via
Feshbach resonance or confinement induced resonance \cite{cir1} for 1D
systems. At half-filling, the Hamiltonian remains invariant under the
simultaneous transformations $\varphi \rightarrow \varphi +\pi $ and $%
c_{j\sigma }\rightarrow c_{j\bar{\sigma}}^{\dag }$, resulting in an energy
spectrum of period-$\pi $ on $\varphi $.

In this paper, we elaborate on $q=3$ case and the generalization to other
periods is straightforward. For incommensurate case, the interaction can
induce many-body localization~\cite{potential3}. Since each unit cell has
three sites, the single-particle spectrum contains three bands [Fig. \ref%
{fig1}(c)]. Different from normal insulator, there are end modes connecting
adjacent bands in each gap under open boundary condition (OBC). Under
periodical boundary condition (PBC), these end modes merge into bulk bands
and disappear. The nontrivial band topology is characterized by the
topological invariant Chern number \cite%
{lang,kraus,shiliangzhu,xuzhihao,xuzhihaotopomott,hhpmag1,hhpmag2}
\begin{equation}
\mathcal{C}=\frac{1}{2\pi }\iint d\theta d\varphi F(\theta ,\varphi )
\label{chernnumber}
\end{equation}%
formulated in the 2D parameter space spanned by $(\theta ,\varphi )\in
\lbrack 0,2\pi ]\times \lbrack 0,2\pi ]$, where $\theta $ is introduced by
imposing a twist boundary condition \cite{twist1,twist2} $\Psi _{j+L}=\Psi
_{j}e^{i\theta }$ on the wave function, $F(\theta ,\varphi )=\text{Im}%
(\langle \frac{\partial \Psi }{\partial \varphi }|\frac{\partial \Psi }{%
\partial \theta }\rangle -\langle \frac{\partial \Psi }{\partial \theta }|%
\frac{\partial \Psi }{\partial \varphi }\rangle )$ is the Berry curvature. $%
\mathcal{C}=1$ and $-1$ when the chemical potential lies in the first and
second band gaps, respectively. According to bulk-edge correspondence, the
Chern number is equal to the number of chiral edge states by adiabatically
evolving $\varphi $, as long as the bulk gap remains finite in the whole
process \cite{thouless,CZpumping,wangleipumping,experimentpump}.

\textit{Topological Mott transition and SCS.---}Now we consider an
interacting many-body system ($N_{\uparrow }$,$N_{\downarrow }$), composed
of $N_{\uparrow }$ spin-up and $N_{\downarrow }$ spin-down atoms. Both total
density $\rho =\sum_{j,\sigma }n_{j\sigma }/L$ and magnetization $%
m=\sum_{j}(n_{j\uparrow }-n_{j\downarrow })/L$ are conserved quantities. We
denote the $n$-th lowest eigenenergy and corresponding wave function as $%
E_{n}(N_{\uparrow },N_{\downarrow })$ and $\Psi _{n}(N_{\uparrow
},N_{\downarrow })$, respectively. $n=0$ then refers to the many-body ground
state. In the subsequent DMRG calculations, density-matrix eigenstates are
kept dynamically to ensure the discarded weight less than $10^{-9}$. The
maximum truncation error of the ground state energy is about $10^{-7}$ and
in general $10-15$ sweeps are enough to reach the required precision.

We utilize three different bulk excitation gaps to characterize the
low-energy modes: charge gap $\Delta _{c}=[E_{0}(N_{\uparrow
}+1,N_{\downarrow }+1)+E_{0}(N_{\uparrow }-1,N_{\downarrow
}-1)-2E_{0}(N_{\uparrow },N_{\downarrow })]/2$ with a fixed magnetization $m$%
, spin gap $\Delta _{s}=[E_{0}(N_{\uparrow }+1,N_{\downarrow
}-1)+E_{0}(N_{\uparrow }-1,N_{\downarrow }+1)-2E_{0}(N_{\uparrow
},N_{\downarrow })]/2$ with a fixed density $\rho $, and neutral gap $\Delta
_{ne}=E_{1}(N_{\uparrow },N_{\downarrow })-E_{0}(N_{\uparrow },N_{\downarrow
})$. The neutral gap directly gives the lowest excitation energy of the
many-body system ($N_{\uparrow }$,$N_{\downarrow }$).

\begin{figure}[t]
\centering
\includegraphics[width=3.2in]{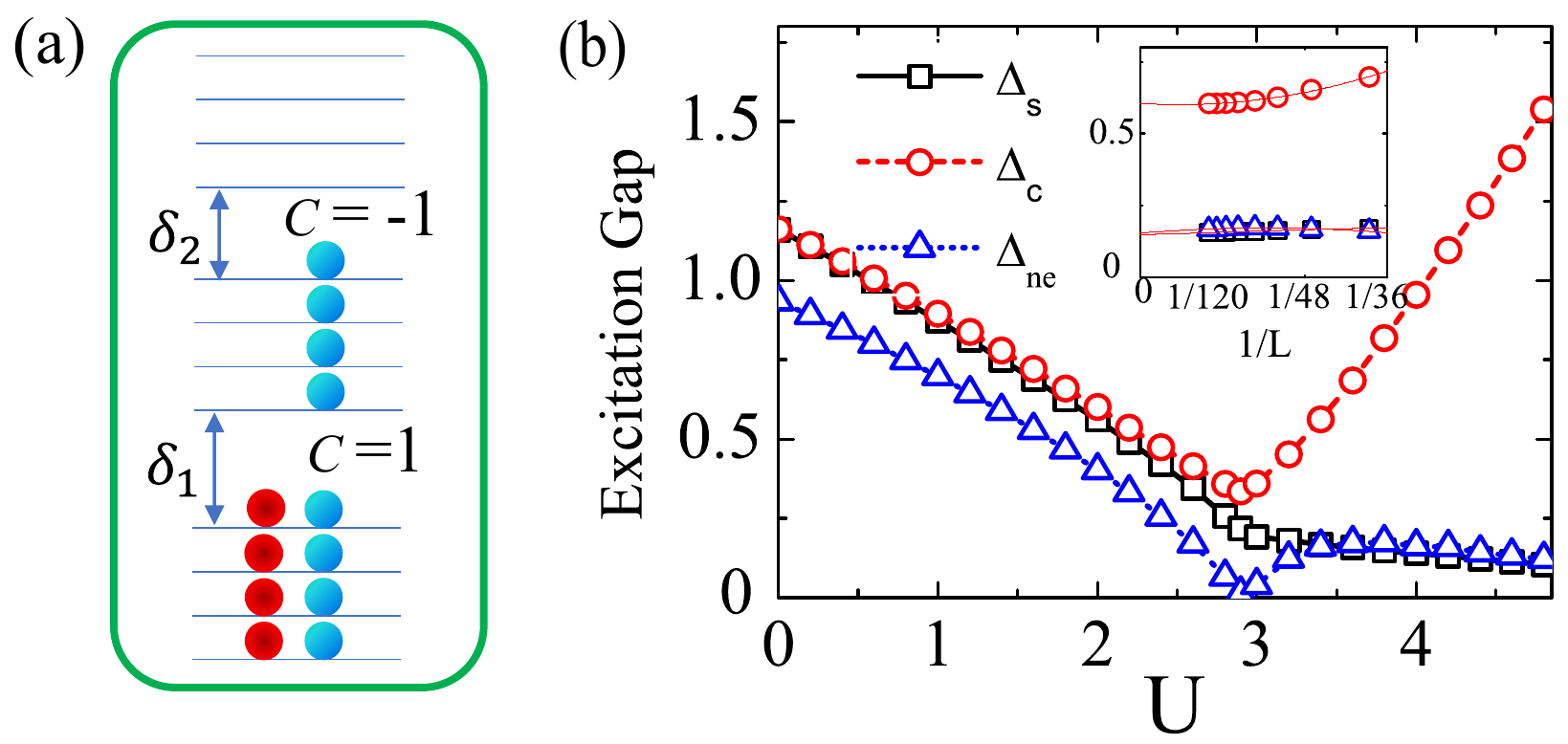}
\caption{(a) Sketch of level occupation in non-interacting limit with $%
\protect\rho =1$, $m=-1/3$. $\protect\delta _{1}$ and $\protect\delta _{2}$
denote the single-particle band gaps, characterized by $\mathcal{C}=\pm 1$,
respectively. (b) The spin gap $\Delta _{s}$ (black square), charge gap $%
\Delta _{c}$ (red circle), and neutral gap $\Delta _{ne}$ (blue triangle)
with respect to interaction $U$. The inset shows finite-size analysis of
three gaps. The red lines are from polynomial fitting. $U=3.5$, $\protect\mu %
=1.2$, $\protect\varphi =0$.}
\label{fig2}
\end{figure}

We concentrate on the half-filling case with $\rho =1$, $m=-1/3$. In the
non-interacting limit, two-component fermionic atoms populate
single-particle levels from low to high independently as sketched in Fig. %
\ref{fig2}(a). The spin-up atoms are filled up to the first band gap
(denoted as $\delta _{1}$) and spin-down atoms to the second band gap
(denoted as $\delta _{2}$). Because two band gaps are characterized by
opposite Chern numbers $\mathcal{C}=\pm 1$, the system is in a quantum spin
Hall-like state with a spin Chern numbers $\mathcal{C}_{s}=2$~\cite%
{spincn1,spincn2} that is defined by Eq. (\ref{chernnumber}) through
choosing $\theta _{\uparrow }=-\theta _{\downarrow }$ for the twisted
boundary condition. Here a single-particle excitation should overcome the
band gaps and carry both charge and spin degree of freedom. The spin and
charge gaps are equal, determined by: $\Delta _{s}=\Delta _{c}=(\delta
_{1}+\delta _{2})/2$ because spin and charge modes are tightly bound
together. $\Delta _{ne}=\min \{\delta _{1},\delta _{2}\}$ in this case.

With increasing interaction, the spin and charge modes would exhibit totally
different behaviors. Fig. \ref{fig2}(b) plots the three gaps with respect to
$U$. Due to the repulsion between different components, the atoms prefer
occupying higher levels. Starting from the same value, both $\Delta _{c}$
and $\Delta _{s}$ decrease first. $\Delta _{c}$ arrives its minimum at $%
U_{c}\approx 2.9$, where $\Delta _{ne}$ closes. $\Delta _{c}$ and $\Delta
_{s}$ coincide with each other at $U<U_{c}$, revealing that spin and charge
modes are coupled together. After the critical point, an obvious separation
between spin and charge excitations happens. $\Delta _{c}$ grows rapidly
(linearly) by further increasing $U$, whereas $\Delta _{s}$ is suppressed
(with $1/U$). Note that none of the three excitation gaps can close for any $%
U>U_{c}$, and the system enters into a Mott insulator phase, with every
lattice sites being populated at $U\rightarrow \infty $. Since the
low-energy excitation now possesses only spin degree of freedom, $\Delta
_{ne}$ coincides with $\Delta _{s}$ in this regime, as clearly demonstrated
by our numerical results. To rule out the size effect, we do a finite-size
scaling of different gaps after Mott transition. As shown in the inset of
Fig. \ref{fig2}(b), $\Delta _{c}$ and $\Delta _{s}$ ($\Delta _{ne}$) both
tend to finite values in thermodynamic limit, which is crucial for the
protection of non-trivial topological properties.

The Mott transition at $U=U_{c}$ is of first-order, with gap closing between
ground and first excited states. Lattice translation symmetry forces~$%
E_{n}(\varphi )=E_{n}(\varphi +\pi /3)$ at $\rho =1$, therefore there are
six gap closing points $\varphi _{p}=p\pi /3$ ($0\leq p\leq 5$ is an
integer) in the whole evolution period of $\varphi \in \lbrack 0,2\pi ]$.
Across the Mott transition, the spin Chern number changes six from $\mathcal{%
C}_{s}=2$ to $\mathcal{C}_{s}=-4$, in consistent with simultaneous gap
closing at six $\varphi _{p}$.

The Mott transition is accompanied with SCS. After the transition, the
Hamiltonian can be represented by two distinct sectors $H=H_{c}+H_{s}$.
While the charge sector $H_{c}$ is fully gapped out due to the strong
interaction, the low-energy physics is governed by an effective spin sector $%
H_{s}$. A second-order perturbation theory at large $U$ leads to an
antiferromagnetic spin superexchange Hamiltonian \cite{supp}
\begin{equation}
H_{s}=\sum_{j}J[1+\frac{(\mu _{j+1}-\mu _{j})^{2}}{U^{2}}]~\mathbf{S}%
_{j}\cdot \mathbf{S}_{j+1}  \label{spinmodel}
\end{equation}%
with periodically modulated exchange couplings. Here $\mathbf{S}%
_{j}=c_{j}^{\dag }{\bm\sigma }c_{j}$/2 is the local spin operator at $j$-th
site. $J=4t^{2}/U$ is the key energy scale of Mott physics. $\Delta _{ne}$ ($%
\Delta _{s}$) decreases by $1/U$ to the leading order. From the standard
bosonization theory \cite{oshikawa}, $m=-1/3$ is a quantized magnetization
plateau that is topologically protected \cite{hhpmag1}. All non-trivial
properties can be understood from the above low-energy theory \cite{supp}.

\textit{Bulk-edge correspondence.---}The appearance of edge states is
usually considered as a hallmark of topological properties. As SCS has
severely changed the nature of low-energy excitations, bulk-edge
correspondence of a topological Mott state is different. To this end, we
study the low-energy spectrum under OBC and demonstrate the consequence of
SCS. The charge distribution of a neutral excitation is $n_{j}^{ne}=n_{j}[%
\Psi _{1}(N_{\uparrow },N_{\downarrow })]-n_{j}[\Psi _{0}(N_{\uparrow
},N_{\downarrow })]$. Here $[\Psi ]$ represents for taking expectation value
on state $\Psi $ and $n_{j}=n_{j\uparrow }+n_{j\downarrow }$. Similarly, the
spin distribution of a neutral excitation is $S_{j}^{ne}=S_{j}^{z}[\Psi
_{1}(N_{\uparrow },N_{\downarrow })]-S_{j}^{z}[\Psi _{0}(N_{\uparrow
},N_{\downarrow })]$ with $S_{j}^{z}=(n_{j\uparrow }-n_{j\downarrow })/2$.
For each end, the accumulated charge (spin) is $n_{l}^{ne}(S_{l}^{ne})=%
\sum_{j<L/2}n_{j}^{ne}(S_{j}^{ne})$, and $n_{r}^{ne}(S_{r}^{ne})=%
\sum_{j>L/2}n_{j}^{ne}(S_{j}^{ne})$, respectively.
\begin{figure}[t]
\centering
\includegraphics[width=3.4in]{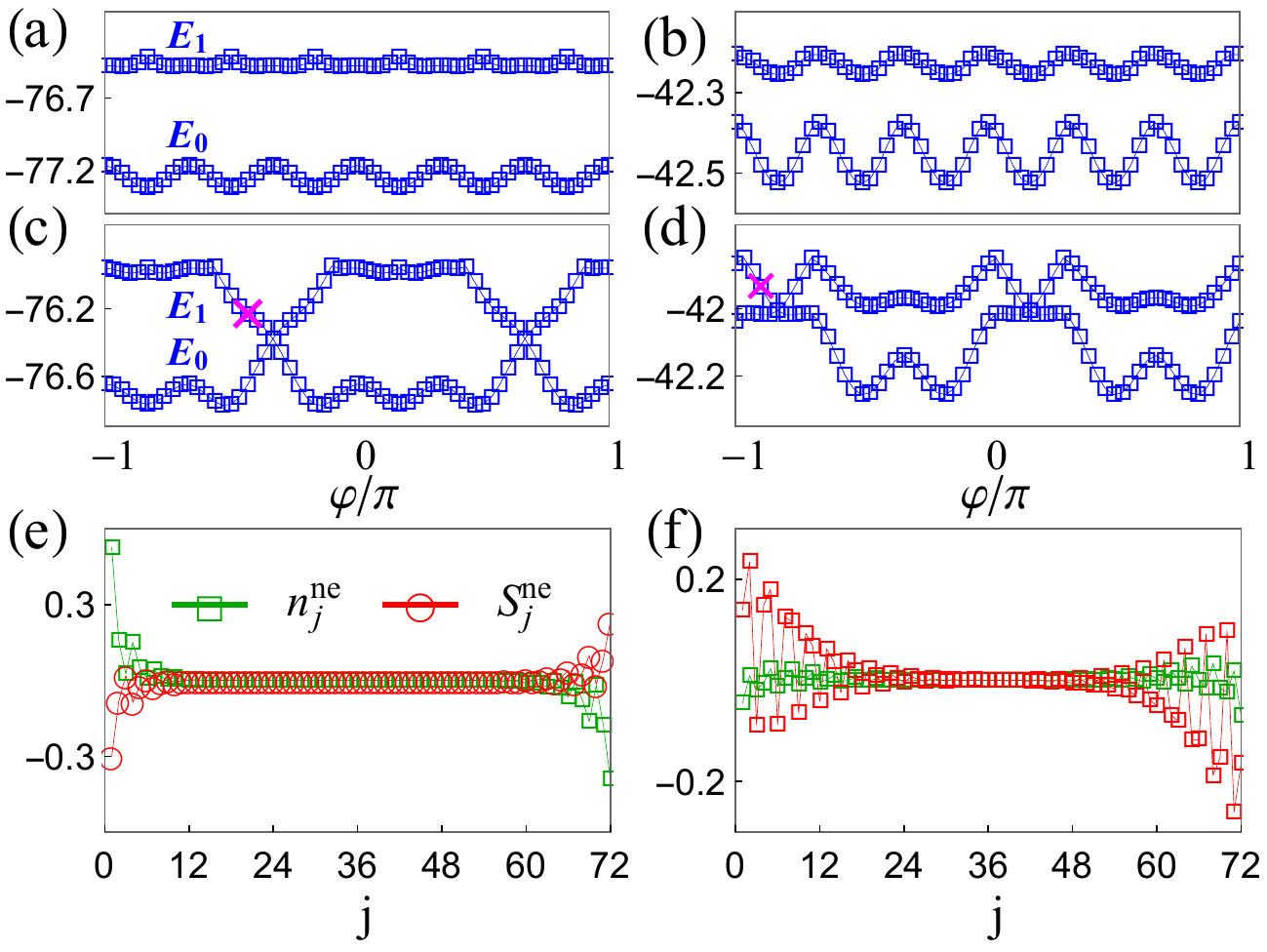}
\caption{The evolution of low-energy spectrum with $\protect\varphi $ for $%
\protect\rho =1$, $m=-1/3$ under (a)~PBC; (c)~OBC for $U=1$ and (b)~PBC;
(d)~OBC for $U=3.5$. (e) and (f) represent the spatial distributions of
charge $n_{j}^{ne}$ (blue square) and spin $S_{j}^{ne}$ (red circle) for the
chosen neutral excitation modes [shown by \textquotedblleft
cross\textquotedblright\ in (c) and (d)], respectively. $U=3.5$. $\protect%
\mu =1.2$.}
\label{fig3}
\end{figure}

Our results are summarized in Fig. \ref{fig3}. Before SCS, there exist
gapless modes inside the neutral gap under OBC by advancing $\varphi $.
While prohibited under PBC [Fig. \ref{fig3}(a)], the ground state and first
excited state cross at $\varphi =-\pi /3,2\pi /3$ under OBC as shown in Fig. %
\ref{fig3}(c), indicating these modes are localized end modes. The crossings
are reminiscent of the level crossings of single-particle levels in Fig. \ref%
{fig1}(c). Fig. \ref{fig3}(e) plots the spatial distributions of spin and
charge for one of the in-gap neutral excitations. It is clear that the
excitation carries both spin and charge at two ends. Our numeric shows $%
n_{l}^{ne}=1$, $S_{l}^{ne}=-1/2$ for the left end and vice versa for the
right end, which is in agreement with the bulk spin Chern number $\mathcal{C}%
_{s}=2$ and validates the single-particle nature of these low-energy modes.

After SCS, the low-energy spectrum is fully gapped under PBC [Fig. \ref{fig3}%
(b)] and gapless end modes still emerge as shown in Fig. \ref{fig3}(d). Note
that the appearance of these modes is at different $\varphi $'s ($\varphi
=-5\pi /6,\pi /6$), which can be understood from the adiabatic continuity
\cite{supp} of the effective\ spin Hamiltonian (\ref{spinmodel}). $%
n_{l,r}^{ne}=0$, and $S_{l,r}^{ne}=\pm 1$ for left and right ends (Fig. \ref%
{fig3}(f)). These two pairs of spin-1 edge modes are in consistent with bulk
$\mathcal{C}_{s}=-4$.
\begin{figure}[t]
\centering
\includegraphics[width=3.3in]{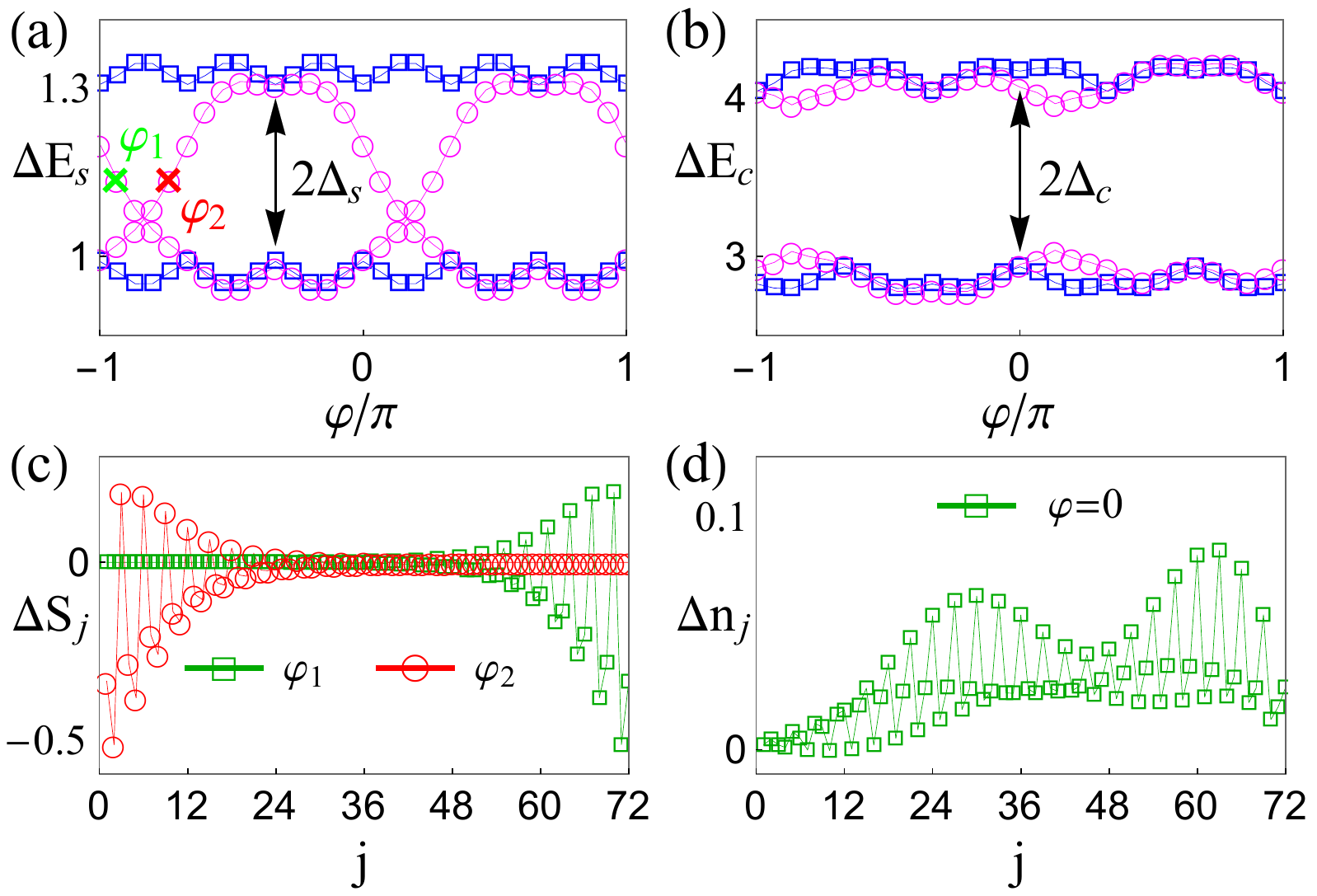}
\caption{(a) Magnon spectrum $\Delta E_{s}(24,48)$ and $\Delta E_{s}(25,47)$
at filling $\protect\rho =1$. The blue and magenta points are for PBC and
OBC, respectively. (b) Charge spectrum $\Delta E_{c}(24,48)$ and $\Delta
E_{c}(23,47)$ for magnetization $m=-1/3$. (c) Spatial distributions $\Delta
S_{j}$ for two in-gap magnon modes [labeled by \textquotedblleft
cross\textquotedblright\ in (a)]. (d) Spatial distributions $\Delta n_{j}$
for the charge excitation at $\protect\varphi =0$. $\protect\mu =1.2$, $%
U=3.5 $. The double-headed arrows denote the spin and charge gaps.}
\label{fig4}
\end{figure}

\textit{Topological magnon excitation.---}The spin-1 low-energy excitations
after SCS indicates the existence of pure collective magnon edge excitations
in the spin sector. Starting from the many-body ground state $\Psi
_{0}(N_{\uparrow },N_{\downarrow })$, the magnon spectrum is defined as $%
\Delta E_{s}(N_{\uparrow },N_{\downarrow })=E_{0}(N_{\uparrow
}-1,N_{\downarrow }+1)-E_{0}(N_{\uparrow },N_{\downarrow })$ by fixing the
total density of the system, with spatial spin distribution $\Delta
S_{j}=S_{j}^{z}[\Psi _{0}(N_{\uparrow }-1,N_{\downarrow }+1)]-S_{j}^{z}[\Psi
_{0}(N_{\uparrow },N_{\downarrow })]$ and no charge. Similarly, we define
charge spectrum with fixed magnetization as $\Delta E_{c}(N_{\uparrow
},N_{\downarrow })=E_{0}(N_{\uparrow }+1,N_{\downarrow
}+1)-E_{0}(N_{\uparrow },N_{\downarrow })$ and associated spatial
distribution as $\Delta n_{j}=n_{j}[\Psi _{0}(N_{\uparrow }+1,N_{\downarrow
}+1)]-n_{j}[\Psi _{0}(N_{\uparrow },N_{\downarrow })]$.

In Fig. \ref{fig4}(a)(b), we show these two spectra after Mott transition,
respectively. A magnon gap ($\sim 2\Delta _{s}$) separates the magnon bands
under PBC, whereas in-gap modes under OBC cross at $\varphi =-5\pi /6$ and $%
\pi /6$ [Fig. \ref{fig4}(a)], same as the neutral excitations in Fig. \ref%
{fig3}(d). The distributions of magnon excitations in two typical phases $%
\varphi =-14\pi /15$, $\varphi =-11\pi /15$ are shown in Fig. \ref{fig4}(c),
which are well localized end modes. Once touching the bulk band, they merge
into the bulk and reappear on the other end. The non-trivial magnon
excitations are closely related to the quantized magnetization plateau \cite%
{hhpmag1} of our effective model (\ref{spinmodel}). For the charge spectrum
in Fig. \ref{fig4}(b), a charge gap ($\sim 2\Delta _{c}$) always exists
regardless of boundary conditions. For any $\varphi $, the charge
excitations distribute on the whole lattice [Fig. \ref{fig4}(d)], revealing
the triviality in the charge sector.

\textit{Discussion and summary.---}Our mechanism of inducing topological
Mott insulator states based on SCS is quite general. The topological Mott
physics here can be extended to i) other period-$q$ (including
incommensurate case, \textit{i.e.}, the famous Aubry-Andr\'{e} model \cite%
{aamodel}); ii) other fillings or magnetizations; iii) off-diagonal
counterparts of model (\ref{ham}), \textit{i.e.}, triple-well lattices with
period modulations on tunneling $t$, instead of on-site energy $\mu _{j}$.
The emergence of topological Mott insulator states with various spin Chern
numbers and spinful edge modes is expected from our low-energy theory.

The proposed topological Mott insulator state and associated low-energy
excitations can be experimentally probed in ultracold atomic gases. In
addition to the proposed scheme using two sets of optical lattices, the 1D
fermionic superlattice can also be generated using the recently developed
digital micromirror device \cite{lattice,lattice2}. For fermionic $^{6}$%
\ce{Li} atoms \cite{hulet,FHAF}, the wavelength of the laser beam is chosen
as $\lambda =1064$ nm, with the recoil energy $E_{r}\approx 2\pi \hbar
\times 29.4$ kHz. At $V_{0}/E_{r}=5$, $t\approx 2\pi \hbar \times 1.9$ kHz.
The neutral and spin gaps of the topological Mott insulator state is then $%
\Delta _{ne}\approx 0.2t=2\pi \hbar \times 390$ Hz, which is large enough
(compared to temperature) to protect the topological properties \cite{FHAF}.
These excitation gaps may be measured using radio-frequency spectroscopy
\cite{gapdetect1,gapdetect2,gapdetect3}. The edge magnon excitations in the
spin sector can be generated using a two-photon Raman process and their spin
distributions at each site can be measured by detecting atomic spin
distributions of different many-body ground states using spin-resolved
quantum gas microscope \cite{detect1,detect2,detect3,detect4,detect5,detect6}
in optical lattices.

In summary, we have studied topological Mott transitions accompanied by SCS in a simple 1D optical superlattices, with low-energy
excitations changing from single-particle spin-1/2 modes to bosonic spin-1
collective modes at the boundary. A novel topological Mott insulator state,
characterized by spin Chern number and gapless magnon excitations is
identified. Our work may pave the way for the experimental observation of
topological Mott insulator states in ultracold atomic gases.

\begin{acknowledgments}
\textbf{Acknowledgments}: We would like to thank Fan Zhang for helpful
discussions. This work is supported by NSF (PHY-1505496),
ARO(W911NF-17-1-0128), AFOSR (FA9550-16-1-0387). S C is supported by the
National Key Research and Development Program of China (2016YFA0300600) and
NSFC under Grants No. 11425419 and No. 11374354.
\end{acknowledgments}

\newpage 
\onecolumngrid
\appendix

\section{Supplementary Materials}

\subsection{Derivation of the effective spin superexchange model}

We give a simple derivation of the effective spin superexchange model (\ref%
{spinmodel}) in the main text using the second-order perturbation theory. To
this end, we split the Hamiltonian (\ref{ham}) into two parts $%
H=Un_{j\uparrow }n_{j\downarrow }+H_{pert}$, with $H_{pert}=\sum_{j,\sigma
}[-t(c_{j\sigma }^{\dag }c_{j+1\sigma }+h.c.)+\mu _{j}n_{j\sigma }]$ as the
perturbation term. For the half-filling case $\rho =1$ and $U\gg t$, the
local Hilbert space on sites $j$ and $j+1$ is spanned by the following four
basis: $|\uparrow _{j},\uparrow _{j+1}\rangle $, $|\uparrow _{j},\downarrow
_{j+1}\rangle $, $|\downarrow _{j},\uparrow _{j+1}\rangle $, $|\downarrow
_{j},\downarrow _{j+1}\rangle $. The effective Hamiltonian can be
represented in this basis as $H_{s}=\sum_{j}H_{j,j+1}$, with
\begin{equation}
H_{j,j+1}=\left(
\begin{array}{cccc}
0 & 0 & 0 & 0 \\
0 & -\frac{t^{2}}{U+\mu _{j+1}-\mu _{j}}-\frac{t^{2}}{U+\mu _{j}-\mu _{j+1}}
& \frac{t^{2}}{U+\mu _{j+1}-\mu _{j}}+\frac{t^{2}}{U+\mu _{j}-\mu _{j+1}} & 0
\\
0 & \frac{t^{2}}{U+\mu _{j+1}-\mu _{j}}+\frac{t^{2}}{U+\mu _{j}-\mu _{j+1}}
& -\frac{t^{2}}{U+\mu _{j+1}-\mu _{j}}-\frac{t^{2}}{U+\mu _{j}-\mu _{j+1}} &
0 \\
0 & 0 & 0 & 0%
\end{array}%
\right) .
\end{equation}%
Using spin-1/2 operator $\mathbf{S}=(S_{x},S_{y},S_{z})$, the above
Hamiltonian can be further represented as
\begin{equation}
H_{s}=\sum_{j}\frac{4t^{2}U}{U^{2}-(\mu _{j+1}-\mu _{j})^{2}}[\mathbf{S}%
_{j}\cdot \mathbf{S}_{j+1}-1/4].
\end{equation}%
By setting $J=\frac{4t^{2}}{U}$ and Taylor expanding the exchange
coefficient by $1/U$ to the second term, we can get the effective spin
superexchange model (\ref{spinmodel}) after neglecting the constant term.

\subsection{Adiabatic continuity and topological properties}

Now we demonstrate the topological properties of the spin superexchange
model, which dictates the low-energy physics of the system. To be more
intuitive and clear, we introduce an anisotropy parameter $g$ in $%
S_{j}^{z}S_{j+1}^{z}$ term:
\begin{equation}
{H}_{s}(g)=\sum_{j}J[1+\frac{(\mu _{j+1}-\mu _{j})^{2}}{U^{2}}%
]~(S_{j}^{x}S_{j+1}^{x}+S_{j}^{y}S_{j+1}^{y}+gS_{j}^{z}S_{j+1}^{z}).
\label{spinmodel2}
\end{equation}%
When $g=1$, the above model recovers our low-energy Hamiltonian (\ref%
{spinmodel}). $g=0$ corresponds to an exactly solvable spin-XX chain by
Jordan-Wigner transformation. We first show the non-trivial topology of $g=0$
case and then demonstrate the adiabatic continuity in the whole region $%
g\geq 0$.

For $g=0$, the Jordan-Wigner transformation (denote $S_{k}^{\pm
}=S_{k}^{x}\pm iS_{k}^{y}$)
\begin{equation*}
d_{j}^{\dag }=e^{i\pi \sum_{k=1}^{j-1}S_{k}^{+}S_{k}^{-}}\cdot
S_{j}^{+},~~~d_{j}=e^{-i\pi \sum_{k=1}^{j-1}S_{k}^{+}S_{k}^{-}}\cdot
S_{j}^{-}
\end{equation*}%
takes the model (\ref{spinmodel2}) to a spinless fermion model: $%
H_{J-W}=\sum_{j}J[1+\frac{(\mu _{j+1}-\mu _{j})^{2}}{U^{2}}](d_{j}^{\dag
}d_{j+1}+d_{j+1}^{\dag }d_{j})$. The band structure is shown in Fig. \ref%
{figs1}(a). Due to the periodically modulated hopping of Jordan-Wigner
fermions, the single-particle spectrum $\varepsilon _{n}^{d}$ consists of
three topological bands and gapless end modes inside the band gap with the
evolution of phase $\varphi $ under OBC. Notice that different from the
original single-particle band in Fig. \ref{fig1}, these end modes now cross
at $\varphi =-5\pi /6$ and $\varphi =\pi /6$ (in consistent with the DMRG
results in Fig. 3(b) and Fig. 4(a) in the main text) and represent for
low-energy collective spinful modes.

\begin{figure}[h]
\centering
\includegraphics[width=6in]{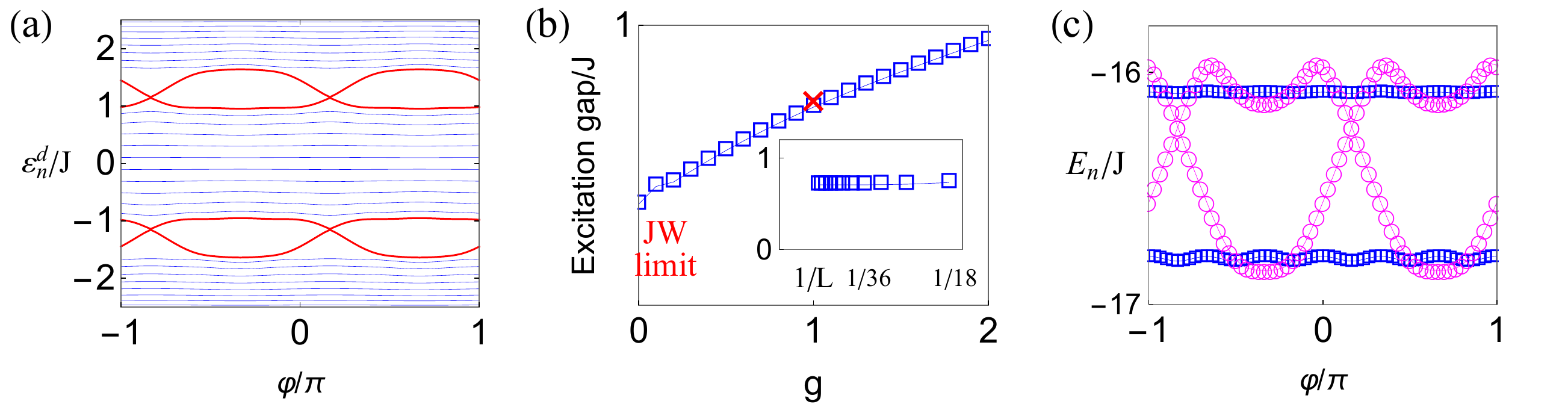}
\caption{(a) Single-particle band of Jordan-Wigner fermions, $g=0$.
(b)~Excitation gap with respect to $g$. The inset shows the finite size
scaling at $g=1$, $\protect\varphi =0$. (c) The lowest two eigenenergies of
model (\protect\ref{spinmodel2}) under OBC (magenta circle) and PBC (blue
square) with the evolution of $\protect\varphi $. $\protect\mu %
^{2}/U^{2}=0.2 $ for all figures.}
\label{figs1}
\end{figure}

To demonstrate the adiabatic continuity in the whole region $g\geq 0$, we
plot the excitation gap (energy difference between the ground state and
first excited state) of model (\ref{spinmodel2}) with respect to $g$ in Fig. %
\ref{figs1}(b). With increasing $g$ from Jordan-Wigner limit $g=0$, the
excitation gap increases. From the scaling behavior of the gap [Inset of
Fig. \ref{figs1}(b)], we can see it tends to a finite value in
thermodynamical limit at $g=1$. The above adiabatic continuity reveals that
our effective spin superexchange model $H_s$ has the same topological
properties with that of Jordan-Wigner fermions.

Furthermore, we show the low-energy spectrum of the spin superexchange model
(\ref{spinmodel}) in Fig. \ref{figs1}(c). Although there always exist an
excitation gap at any $\varphi $ under PBC, the ground state and first
excited state touch at $\varphi =-5\pi /6$ and $\varphi =\pi /6$ under OBC
(in consistent with the DMRG results in Fig. 3(b) and Fig. 4(a) in the main
text). As these crossings depend on boundary condition, the excitations
around the touching points are well localized end modes [not shown]. For
example, at $\varphi =-9\pi /10$, our numeric gives accumulated spin
distribution $\Delta S_{z}=0.9996$ on the left end and $\Delta S_{z}=-0.9996$
on the right end. The integer spinful end modes are in consistent with those
neutral excitations in the main text.

\end{document}